\documentclass[twocolumn,showpacs,amsmath,amssymb,pra]{revtex4}
\usepackage{graphicx}
\usepackage{bm}
\newcommand{\vdW}{r_\textrm{vdW}}

\newcommand{\eq}[1]{Eq.~\eqref{#1}}
\begin{document}

\title{Efimov Physics and the Three-Body Parameter within a Two-Channel Framework}
\author{P.~K. S{\o}rensen}
\author{D.~V. Fedorov}
\author{A.~S. Jensen}
\author{N.~T. Zinner}
\affiliation{Department of Physics and Astronomy, Aarhus University, DK-8000 Aarhus C, Denmark}
\date{\today}

\begin{abstract}
We calculate shallow three-body bound states in the universal regime, defined by Efimov,
with inclusion of both scattering length and effective range parameters. 
We
find corrections to the universal scaling laws for large binding energies.
For narrow resonances we find
a distinct non-monotonic behavior of the threshold at which the lowest Efimov trimer 
merges with the three-body continuum. 
The origin of the three-body parameter is related to
the two-body atom-atom interactions in a physically clear model.
Our results demonstrate that experimental information from narrow Feshbach resonances
and/or mixed systems are of vital importance to pin down the 
relation of two- and three-body physics in atomic systems.
\end{abstract}
\maketitle

\paragraph*{Introduction.}
The counterintuitive behavior of three-body systems at 
the threshold for two-body binding is highlighted by the 
Efimov effect where an infinitude of geometrically scaling 
states appears \cite{efi70}. While unsuccesfully 
sought for in nuclear physics \cite{jen04}, the effect has 
been confirmed and explored in ultracold atomic gases \cite{ferlaino2010}.
From the theoretical point of view these systems have been 
described accurately by universal theories that 
only take the lowest order scattering dynamics into account 
through the two-body scattering length, $a$ \cite{nie01,braaten2006}.
However, the overall scale of the spectrum cannot be obtained
in the universal theory and the so-called three-body parameter, $\Lambda$, 
is needed to complete the formalism.

In ultracold atomic gases, trimer physics can be studied using 
inter-atomic Feshbach resonances \cite{chin2010} that provide
tunability of $a$ over many orders of magnitude. The parameters
of these resonances will in general depend on the microscopic
details of a particular atomic system. Still, in a surprising
development, the Grimm group have reported strong indications
that the three-body parameter is determined by the van der Waals 
length, $r_\textrm{vdW}$ \cite{berninger2011}. This implies
that atomic trimer physics in the weakly bound limit only depend 
on two-body parameters.
The experimental findings have generated a flurry of recent
theoretical interest \cite{naidon2011,chin2011,wang2012,schmidt2012}. 
It has been suggested that the 
presence of many deep bound states in the two-body potential
typical of alkali atom systems will suppress the dependence
on short-range physics due to a
large inner repulsive barrier \cite{chin2011,wang2012}.

\begin{figure}[ht!]
\centering
\begingroup
  \makeatletter
  \providecommand\color[2][]{%
    \GenericError{(gnuplot) \space\space\space\@spaces}{%
      Package color not loaded in conjunction with
      terminal option `colourtext'%
    }{See the gnuplot documentation for explanation.%
    }{Either use 'blacktext' in gnuplot or load the package
      color.sty in LaTeX.}%
    \renewcommand\color[2][]{}%
  }%
  \providecommand\includegraphics[2][]{%
    \GenericError{(gnuplot) \space\space\space\@spaces}{%
      Package graphicx or graphics not loaded%
    }{See the gnuplot documentation for explanation.%
    }{The gnuplot epslatex terminal needs graphicx.sty or graphics.sty.}%
    \renewcommand\includegraphics[2][]{}%
  }%
  \providecommand\rotatebox[2]{#2}%
  \@ifundefined{ifGPcolor}{%
    \newif\ifGPcolor
    \GPcolorfalse
  }{}%
  \@ifundefined{ifGPblacktext}{%
    \newif\ifGPblacktext
    \GPblacktexttrue
  }{}%
  \let\gplgaddtomacro\g@addto@macro
  \gdef\gplbacktext{}%
  \gdef\gplfronttext{}%
  \makeatother
  \ifGPblacktext
    \def\colorrgb#1{}%
    \def\colorgray#1{}%
  \else
    \ifGPcolor
      \def\colorrgb#1{\color[rgb]{#1}}%
      \def\colorgray#1{\color[gray]{#1}}%
      \expandafter\def\csname LTw\endcsname{\color{white}}%
      \expandafter\def\csname LTb\endcsname{\color{black}}%
      \expandafter\def\csname LTa\endcsname{\color{black}}%
      \expandafter\def\csname LT0\endcsname{\color[rgb]{1,0,0}}%
      \expandafter\def\csname LT1\endcsname{\color[rgb]{0,1,0}}%
      \expandafter\def\csname LT2\endcsname{\color[rgb]{0,0,1}}%
      \expandafter\def\csname LT3\endcsname{\color[rgb]{1,0,1}}%
      \expandafter\def\csname LT4\endcsname{\color[rgb]{0,1,1}}%
      \expandafter\def\csname LT5\endcsname{\color[rgb]{1,1,0}}%
      \expandafter\def\csname LT6\endcsname{\color[rgb]{0,0,0}}%
      \expandafter\def\csname LT7\endcsname{\color[rgb]{1,0.3,0}}%
      \expandafter\def\csname LT8\endcsname{\color[rgb]{0.5,0.5,0.5}}%
    \else
      \def\colorrgb#1{\color{black}}%
      \def\colorgray#1{\color[gray]{#1}}%
      \expandafter\def\csname LTw\endcsname{\color{white}}%
      \expandafter\def\csname LTb\endcsname{\color{black}}%
      \expandafter\def\csname LTa\endcsname{\color{black}}%
      \expandafter\def\csname LT0\endcsname{\color{black}}%
      \expandafter\def\csname LT1\endcsname{\color{black}}%
      \expandafter\def\csname LT2\endcsname{\color{black}}%
      \expandafter\def\csname LT3\endcsname{\color{black}}%
      \expandafter\def\csname LT4\endcsname{\color{black}}%
      \expandafter\def\csname LT5\endcsname{\color{black}}%
      \expandafter\def\csname LT6\endcsname{\color{black}}%
      \expandafter\def\csname LT7\endcsname{\color{black}}%
      \expandafter\def\csname LT8\endcsname{\color{black}}%
    \fi
  \fi
  \setlength{\unitlength}{0.0500bp}%
  \begin{picture}(5040.00,3528.00)%
    \gplgaddtomacro\gplbacktext{%
      \csname LTb\endcsname%
      \put(946,887){\makebox(0,0)[r]{\strut{}-0.6}}%
      \put(946,1252){\makebox(0,0)[r]{\strut{}-0.5}}%
      \put(946,1618){\makebox(0,0)[r]{\strut{}-0.4}}%
      \put(946,1983){\makebox(0,0)[r]{\strut{}-0.3}}%
      \put(946,2349){\makebox(0,0)[r]{\strut{}-0.2}}%
      \put(946,2715){\makebox(0,0)[r]{\strut{}-0.1}}%
      \put(946,3080){\makebox(0,0)[r]{\strut{} 0}}%
      \put(1078,484){\makebox(0,0){\strut{}-2}}%
      \put(1726,484){\makebox(0,0){\strut{}-1}}%
      \put(2374,484){\makebox(0,0){\strut{} 0}}%
      \put(3023,484){\makebox(0,0){\strut{} 1}}%
      \put(3671,484){\makebox(0,0){\strut{} 2}}%
      \put(4319,484){\makebox(0,0){\strut{} 3}}%
      \put(176,1983){\rotatebox{-270}{\makebox(0,0){\strut{}$\vdW/a^{(-)}$}}}%
      \put(2860,154){\makebox(0,0){\strut{}$\log(s_\text{res})$}}%
      \put(2115,3080){\makebox(0,0)[l]{\strut{}${}^7$Li}}%
      \put(2504,3080){\makebox(0,0)[l]{\strut{}${}^{39}$K}}%
      \put(3087,3080){\makebox(0,0)[l]{\strut{}${}^{85}$Rb}}%
      \put(3865,3080){\makebox(0,0)[l]{\strut{}${}^{133}$Cs}}%
    }%
    \gplgaddtomacro\gplfronttext{%
      \csname LTb\endcsname%
      \put(2600,1097){\makebox(0,0)[l]{\strut{}One-channel}}%
      \csname LTb\endcsname%
      \put(2600,877){\makebox(0,0)[l]{\strut{}Two-channel}}%
    }%
    \gplbacktext
    \put(0,0){\includegraphics{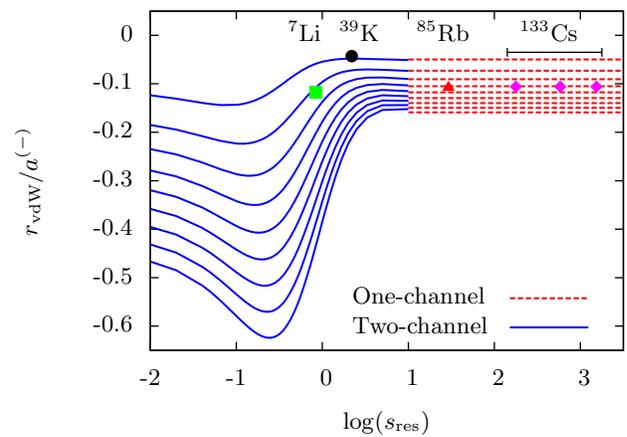}}
    \gplfronttext
  \end{picture}%
\endgroup
\caption{(Color online) The threshold scattering length $a^{(-)}$ at which the lowest
universal Efimov trimer merges with the three-atom continuum plotted against 
the strength, $s_\text{res}$, of the Feshbach resonance. The right-hand side 
corresponds to broad resonances. The different curves of like linetype 
show result with different three-body parameters, $\rho_c$, in units
of the atomic van der Waals length, $r_\textrm{vdW}$. The values 
of $\rho_c$ decrease from top to bottom (precise values are given in
the text). Experimental values are from $^{133}$Cs \cite{kraemer2006,berninger2011}, $^7$Li \cite{pollack2009}, 
$^{39}$K \cite{zaccanti2009}, and $^{85}$Rb \cite{wild2012}.}
\label{fig1}
\end{figure}

The most natural way to eliminate any need for a three-body
parameter is to use a finite-range two-body potential that
exhibits the known features of the inter-atomic interaction.
Here we pursue this natural and straightforward approach 
to the issue of the 
origin of $\Lambda$ in three-body physics. As three-body 
calculations with finite-range potentials are typically 
very cumbersome, it is highly desirable to have simple
models that mimic the finite-range physics. Therefore we 
consider here a model with zero-range interactions that
includes finite-range correction effects.
Consequently, we solve the three-body 
Schr{\"o}dinger equation using
one- and two-channel models for the description of 
the Feshbach resonances that are used to control interactions
in experiments \cite{chin2010}. The three-body parameter
is considered parametrically in both models, and this allows
us to study the threshold behavior of Efimov trimers 
as function of $\Lambda$ and the width of the Fesbhach resonance
which is related to the effective range, $r_{e}$ \cite{chin2010}. 
Here $r_\textrm{vdW}$ and $r_e$ are intimately connected through 
semiclassical calculations \cite{PhysRevA.48.1993,gao1998,PhysRevA.59.1998}. It
is conceivable that this explains the observed values of $\Lambda$ 
as $r_e$ or $r_\textrm{vdW}$ provides a background scale 
that determines the overall scale. However, given the delicate 
non-classical nature of the universal
trimer states it is far from obvious if and how this can work 
out.

The experimental findings indicate that for several different atomic species
the ratio of threshold scattering length $a^{(-)}$ for creation of Efimov
trimers out of the three-body continuum is of 
order $a^{(-)}/r_\textrm{vdW}=8.5\sim9.5$ \cite{chin2011}. Here we show 
that this result can be obtained in a single-channel zero-range model 
for a specific choice of $\Lambda$, which
can be related to the underlying two-body atom-atom physics in a natural way. 
Within a two-channel model we find an intriguing non-trivial behavior
of $a^{(-)}$ for narrow resonances, {\it irrespective} of the three-body
parameter, see Fig.~\ref{fig1}. We also study the behavior of $a^{(-)}$ as a function of 
the number of bound states allowed by the two-body atomic potential for
both one- and two-channel models. Generally, we find that the 
inclusion of effective range decreases $|a^{(-)}|$. 
Our results 
predict that narrow resonance systems are important for
obtaining a full picture of the relation between two- and 
three-body parameters for universal bound state physics.

\paragraph*{Method.}
We consider a system of three identical bosonic particles using hyper\-spherical 
coordinates \cite{suzuki1998} defined from the Cartesian coordinates $\bm r_i$,$\bm r_j$,$\bm r_k$ 
through $\bm x_i=(\bm r_j-\bm r_k)/\sqrt{2}$ and 
$\bm y_i=\tfrac{2}{3}(\bm r_i-(\bm r_j+\bm r_k)/2)$ as
hyperradius, $\rho=\sqrt{\bm x_{i}^{2}+\bm y_{i}^{2}}$, and hyperangle, $\alpha_i=\tan^{-1}\tfrac{|\bm x_i|}{|\bm y_i|}$.
$\{i,j,k\}$ are cyclic permutations of $\{1,2,3\}$ and $\rho$ is independent of this choice.
We apply the hyperspherical adiabatic approach with wave function
$\Psi(\rho,\Omega)=\rho^{-5/2}\sum_nf_n(\rho)\Phi_n(\rho,\Omega)$,
where $\Omega=\{\alpha_i,\bm x_i/|\bm x_i|,\bm y_i/|\bm y_i| \}$ is a set of angular coordinates.
We keep only the lowest adiabatic potential corresponding to $n=0$, 
and index $n$ is henceforth suppressed. 
For the description of Efimov trimer states, 
this approximation has proven extremely accurate
\cite{nie01}. The radial equation is
\begin{equation}
	\left(-\frac{d^2}{d\rho^2}+\frac{\lambda(\rho)+15/4}{\rho^2}-\frac{2mE}{\hbar^2}\right)f(\rho)=0\;,
	\label{eq:4}
\end{equation}
where $\lambda(\rho)$ is the eigenvalue to the hyperangular equation
\begin{equation}
	\left(\tilde \Lambda^2+\frac{2m\rho^2}{\hbar^2}V\right)\Phi(\rho,\Omega)=\lambda(\rho)\Phi(\rho,\Omega),
	\label{eq:5}
\end{equation}
in which $\tilde\Lambda^2$ is the generalized angular momentum operator, $V$ is the two-particle 
interaction potentials, and $m$ is the atomic mass. 
In Eq.~\eqref{eq:4}, the non-adiabatic
corrections are omitted as they are found to be negligible. 

\begin{figure}[ht!]
\centering
\begingroup
  \makeatletter
  \providecommand\color[2][]{%
    \GenericError{(gnuplot) \space\space\space\@spaces}{%
      Package color not loaded in conjunction with
      terminal option `colourtext'%
    }{See the gnuplot documentation for explanation.%
    }{Either use 'blacktext' in gnuplot or load the package
      color.sty in LaTeX.}%
    \renewcommand\color[2][]{}%
  }%
  \providecommand\includegraphics[2][]{%
    \GenericError{(gnuplot) \space\space\space\@spaces}{%
      Package graphicx or graphics not loaded%
    }{See the gnuplot documentation for explanation.%
    }{The gnuplot epslatex terminal needs graphicx.sty or graphics.sty.}%
    \renewcommand\includegraphics[2][]{}%
  }%
  \providecommand\rotatebox[2]{#2}%
  \@ifundefined{ifGPcolor}{%
    \newif\ifGPcolor
    \GPcolorfalse
  }{}%
  \@ifundefined{ifGPblacktext}{%
    \newif\ifGPblacktext
    \GPblacktexttrue
  }{}%
  \let\gplgaddtomacro\g@addto@macro
  \gdef\gplbacktext{}%
  \gdef\gplfronttext{}%
  \makeatother
  \ifGPblacktext
    \def\colorrgb#1{}%
    \def\colorgray#1{}%
  \else
    \ifGPcolor
      \def\colorrgb#1{\color[rgb]{#1}}%
      \def\colorgray#1{\color[gray]{#1}}%
      \expandafter\def\csname LTw\endcsname{\color{white}}%
      \expandafter\def\csname LTb\endcsname{\color{black}}%
      \expandafter\def\csname LTa\endcsname{\color{black}}%
      \expandafter\def\csname LT0\endcsname{\color[rgb]{1,0,0}}%
      \expandafter\def\csname LT1\endcsname{\color[rgb]{0,1,0}}%
      \expandafter\def\csname LT2\endcsname{\color[rgb]{0,0,1}}%
      \expandafter\def\csname LT3\endcsname{\color[rgb]{1,0,1}}%
      \expandafter\def\csname LT4\endcsname{\color[rgb]{0,1,1}}%
      \expandafter\def\csname LT5\endcsname{\color[rgb]{1,1,0}}%
      \expandafter\def\csname LT6\endcsname{\color[rgb]{0,0,0}}%
      \expandafter\def\csname LT7\endcsname{\color[rgb]{1,0.3,0}}%
      \expandafter\def\csname LT8\endcsname{\color[rgb]{0.5,0.5,0.5}}%
    \else
      \def\colorrgb#1{\color{black}}%
      \def\colorgray#1{\color[gray]{#1}}%
      \expandafter\def\csname LTw\endcsname{\color{white}}%
      \expandafter\def\csname LTb\endcsname{\color{black}}%
      \expandafter\def\csname LTa\endcsname{\color{black}}%
      \expandafter\def\csname LT0\endcsname{\color{black}}%
      \expandafter\def\csname LT1\endcsname{\color{black}}%
      \expandafter\def\csname LT2\endcsname{\color{black}}%
      \expandafter\def\csname LT3\endcsname{\color{black}}%
      \expandafter\def\csname LT4\endcsname{\color{black}}%
      \expandafter\def\csname LT5\endcsname{\color{black}}%
      \expandafter\def\csname LT6\endcsname{\color{black}}%
      \expandafter\def\csname LT7\endcsname{\color{black}}%
      \expandafter\def\csname LT8\endcsname{\color{black}}%
    \fi
  \fi
  \setlength{\unitlength}{0.0500bp}%
  \begin{picture}(5040.00,3276.00)%
    \gplgaddtomacro\gplbacktext{%
      \csname LTb\endcsname%
      \put(946,704){\makebox(0,0)[r]{\strut{}-1}}%
      \put(946,1165){\makebox(0,0)[r]{\strut{}-0.8}}%
      \put(946,1627){\makebox(0,0)[r]{\strut{}-0.6}}%
      \put(946,2088){\makebox(0,0)[r]{\strut{}-0.4}}%
      \put(946,2550){\makebox(0,0)[r]{\strut{}-0.2}}%
      \put(946,3011){\makebox(0,0)[r]{\strut{} 0}}%
      \put(1078,484){\makebox(0,0){\strut{}-1}}%
      \put(1969,484){\makebox(0,0){\strut{}-0.5}}%
      \put(2861,484){\makebox(0,0){\strut{} 0}}%
      \put(3752,484){\makebox(0,0){\strut{} 0.5}}%
      \put(4643,484){\makebox(0,0){\strut{} 1}}%
      \put(176,1857){\rotatebox{-270}{\makebox(0,0){\strut{}$\frac{2mr_\textrm{vdW}^{2}}{\hbar^2}E_T$}}}%
      \put(2860,154){\makebox(0,0){\strut{}$\textrm{sign}(a)\times\frac{r_\textrm{vdW}^{2}}{a^2}$}}%
      \put(1212,1235){\makebox(0,0)[l]{\strut{}$a^{(-)}$}}%
    }%
    \gplgaddtomacro\gplfronttext{%
      \csname LTb\endcsname%
      \put(1905,1097){\makebox(0,0)[l]{\strut{}One-channel}}%
      \csname LTb\endcsname%
      \put(1905,877){\makebox(0,0)[l]{\strut{}Two-channel}}%
    }%
    \gplbacktext
    \put(0,0){\includegraphics{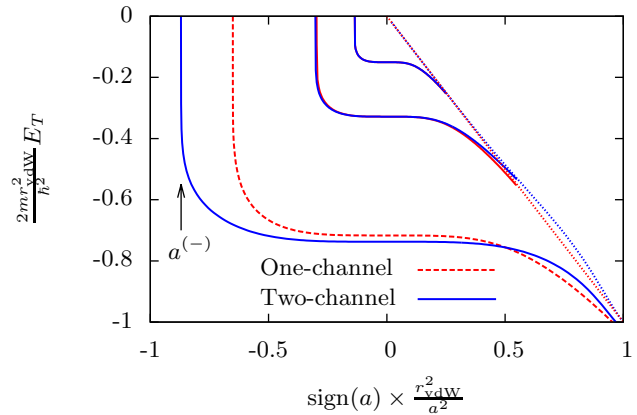}}
    \gplfronttext
  \end{picture}%
\endgroup
\caption{(Color online) Trimer binding energy $E_T$ vs. scattering length $a$ for the one-
and two-channel models with effective range $r_e=-5r_\textrm{vdW}$.
$a^{(-)}$ is the threshold scattering length for 
existence of the lowest trimer. Finely dotted lines indicate atom-dimer threshold for 
positive $a$. Successive states have a scaling of 515 \cite{efi70} and 
for visibility the axes are scaled by the power $1/8$.}
\label{fig2}
\end{figure}

We use zero-range single-channel and a two-channel 
interaction models \cite{FewBody2012}. The former is accurate for broad Feshbach resonances
(small $r_e$) and the latter for
narrow resonances (large $r_e$).
Both models contain the background scattering length in the open 
channel, $a_\textrm{open}$. Our model must predict three-body properties solely
from two-body potentials, so we must insist that $a_\textrm{open}$ is a 
quantity most naturally associated to the two-body $r_\textrm{vdW}$. 
In a semiclassical approach, 
the scattering length of such potentials is 
$\bar{a}\equiv 4\pi (\Gamma[1/4])^{-2}r_\textrm{vdW}\approx 0.956r_\textrm{vdW}$ \cite{PhysRevA.48.1993}.
Then $a_\textrm{open}$ should be identified with $r_\textrm{vdW}$
to within a few percent. The resonance strength 
is related to $r_e$ \cite{FewBody2012}.
We characterize the strength by 
$s_\textrm{res}=r_\textrm{vdW}/|R_0|$, 
where $R_0$ is the effective range when $|a|=\infty$ \cite{chin2010}.

Eq.~\eqref{eq:4} is solved numerically for $f(\rho)$ with the condition 
$f(\rho_\text{L})=0$ for large $\rho_\text{L}$. 
At short distance the zero-range models used here require a cut-off, $\rho_c$, 
with $f(\rho_c)=0$ \cite{nie01}. 
Initially, we consider $\rho_c$ a parameter and study the 
three-body spectrum for different $\rho_c$ and different $s_\textrm{res}$.
$\rho_c$ is the coordinate-space equivalent of $\Lambda$. 
Below we relate the $\rho_c$ to the two-body atomic potential.

\paragraph*{Results.}
In Fig.~\ref{fig2} we show trimer energies, $E_T$, as function of $a$
for both one- and two-channel models. We find that the two-channel energies
are generally lowest when $a<0$ and for $a>0$ the trimers cross
before merging with the atom-dimer 
continuum. As indicated in Fig.~\ref{fig2}, $a^{(-)}$ is the threshold 
scattering length for appearance of the lowest Efimov trimer 
on the negative $a$ side. The two-channel model is here seen to move 
this threshold to the left, i.e. to smaller negative $a^{(-)}$. 
These thresholds and 
the energy for $|a|=\infty$ are connected by universal relations, 
where lower energy on resonance translates to smaller 
$|a^{(-)}|$ at the threshold \cite{efi70}.

A systematic study of the influence of both $\rho_c$ and $r_e$ 
(or equivalently $s_\textrm{res}$) is shown in Fig.~\ref{fig1} which 
is one of our main results. The values of $\rho_c/r_\textrm{vdW}$
from top to bottom in Fig.~\ref{fig1} are 1.20, 0.82, 0.66, 0.58, 
0.51, 0.47, 0.42, 0.40, and 0.38. They correspond to $n=0$ to 9 
in Eq.~\eqref{eq:15} below. 
Both models
agree for $s_\textrm{res}\gg 1$ and we plot the two-channel model results
only in the region where it deviates.
We find that to reproduce the 
experimental data for $s_\textrm{res}\gg 1$, a cut-off of 
$\rho_c/r_\textrm{vdW}=0.58$ is required. However, for small
$s_\textrm{res}$, the same cut-off does not reproduce the 
known data point coming from $^{7}$Li (other measurements 
have smaller $a^{(-)}$ \cite{gross2009,gross2010}, which is closer
to our predictions). While we do find an 
increase toward the $^{39}$K data point at small $s_\textrm{res}$, 
it cannot be accomodated for the same $\rho_c$.

In general, we find that, irrespective of $\rho_c$, the inclusion 
of effective range brings a non-monotonic behavior to $a^{(-)}$, 
and it tends to push the value of  $a^{(-)}$ {\it down} for small 
$s_\textrm{res}$ or large $|r_e|$. The reason can be seen in Fig.~\ref{fig6} 
where the adiabatic potentials for different effective 
ranges, $r_e/r_\textrm{vdw} = 0, -0.1, -1, -5$, are shown. 
At $r_e=0$ the bound states generally reside at large $\rho/r_\textrm{vdW}$ 
\cite{nie01}. The effective range causes an additional repulsive 
barrier initially leading to less bound energies when $|r_e|$ is 
small, since the bound state sitting at large $\rho$ will feel
the barrier at first. When the effective range increases the 
attractive region will 
ultimately lead to an increase in bound state energy as the 
wave function starts to occupy the pocket at smaller $\rho$.

This is in sharp contrast to the study in Ref.~\cite{schmidt2012}
which finds the opposite behavior. 
As argued above, it is physically
reasonable since $E_T$ is lower at finite $r_e<0$. 
We notice that our two-channel model has 
$r_e<0$ which is consistent with the usual theory of Feshbach 
resonances \cite{chin2010}. Ref.~\cite{schmidt2012}
appears to accomodate also $r_e>0$ and this may 
resolve the discrepancy.

\begin{figure}[ht!]
\begingroup
  \makeatletter
  \providecommand\color[2][]{%
    \GenericError{(gnuplot) \space\space\space\@spaces}{%
      Package color not loaded in conjunction with
      terminal option `colourtext'%
    }{See the gnuplot documentation for explanation.%
    }{Either use 'blacktext' in gnuplot or load the package
      color.sty in LaTeX.}%
    \renewcommand\color[2][]{}%
  }%
  \providecommand\includegraphics[2][]{%
    \GenericError{(gnuplot) \space\space\space\@spaces}{%
      Package graphicx or graphics not loaded%
    }{See the gnuplot documentation for explanation.%
    }{The gnuplot epslatex terminal needs graphicx.sty or graphics.sty.}%
    \renewcommand\includegraphics[2][]{}%
  }%
  \providecommand\rotatebox[2]{#2}%
  \@ifundefined{ifGPcolor}{%
    \newif\ifGPcolor
    \GPcolorfalse
  }{}%
  \@ifundefined{ifGPblacktext}{%
    \newif\ifGPblacktext
    \GPblacktexttrue
  }{}%
  \let\gplgaddtomacro\g@addto@macro
  \gdef\gplbacktext{}%
  \gdef\gplfronttext{}%
  \makeatother
  \ifGPblacktext
    \def\colorrgb#1{}%
    \def\colorgray#1{}%
  \else
    \ifGPcolor
      \def\colorrgb#1{\color[rgb]{#1}}%
      \def\colorgray#1{\color[gray]{#1}}%
      \expandafter\def\csname LTw\endcsname{\color{white}}%
      \expandafter\def\csname LTb\endcsname{\color{black}}%
      \expandafter\def\csname LTa\endcsname{\color{black}}%
      \expandafter\def\csname LT0\endcsname{\color[rgb]{1,0,0}}%
      \expandafter\def\csname LT1\endcsname{\color[rgb]{0,1,0}}%
      \expandafter\def\csname LT2\endcsname{\color[rgb]{0,0,1}}%
      \expandafter\def\csname LT3\endcsname{\color[rgb]{1,0,1}}%
      \expandafter\def\csname LT4\endcsname{\color[rgb]{0,1,1}}%
      \expandafter\def\csname LT5\endcsname{\color[rgb]{1,1,0}}%
      \expandafter\def\csname LT6\endcsname{\color[rgb]{0,0,0}}%
      \expandafter\def\csname LT7\endcsname{\color[rgb]{1,0.3,0}}%
      \expandafter\def\csname LT8\endcsname{\color[rgb]{0.5,0.5,0.5}}%
    \else
      \def\colorrgb#1{\color{black}}%
      \def\colorgray#1{\color[gray]{#1}}%
      \expandafter\def\csname LTw\endcsname{\color{white}}%
      \expandafter\def\csname LTb\endcsname{\color{black}}%
      \expandafter\def\csname LTa\endcsname{\color{black}}%
      \expandafter\def\csname LT0\endcsname{\color{black}}%
      \expandafter\def\csname LT1\endcsname{\color{black}}%
      \expandafter\def\csname LT2\endcsname{\color{black}}%
      \expandafter\def\csname LT3\endcsname{\color{black}}%
      \expandafter\def\csname LT4\endcsname{\color{black}}%
      \expandafter\def\csname LT5\endcsname{\color{black}}%
      \expandafter\def\csname LT6\endcsname{\color{black}}%
      \expandafter\def\csname LT7\endcsname{\color{black}}%
      \expandafter\def\csname LT8\endcsname{\color{black}}%
    \fi
  \fi
  \setlength{\unitlength}{0.0500bp}%
  \begin{picture}(5040.00,3528.00)%
    \gplgaddtomacro\gplbacktext{%
      \csname LTb\endcsname%
      \put(946,704){\makebox(0,0)[r]{\strut{}-5.4}}%
      \put(946,988){\makebox(0,0)[r]{\strut{}-5.3}}%
      \put(946,1273){\makebox(0,0)[r]{\strut{}-5.2}}%
      \put(946,1557){\makebox(0,0)[r]{\strut{}-5.1}}%
      \put(946,1841){\makebox(0,0)[r]{\strut{}-5}}%
      \put(946,2126){\makebox(0,0)[r]{\strut{}-4.9}}%
      \put(946,2410){\makebox(0,0)[r]{\strut{}-4.8}}%
      \put(946,2694){\makebox(0,0)[r]{\strut{}-4.7}}%
      \put(946,2979){\makebox(0,0)[r]{\strut{}-4.6}}%
      \put(946,3263){\makebox(0,0)[r]{\strut{}-4.5}}%
      \put(1078,484){\makebox(0,0){\strut{} 0}}%
      \put(1672,484){\makebox(0,0){\strut{} 5}}%
      \put(2266,484){\makebox(0,0){\strut{} 10}}%
      \put(2861,484){\makebox(0,0){\strut{} 15}}%
      \put(3455,484){\makebox(0,0){\strut{} 20}}%
      \put(4049,484){\makebox(0,0){\strut{} 25}}%
      \put(4643,484){\makebox(0,0){\strut{} 30}}%
      \put(176,1983){\rotatebox{-270}{\makebox(0,0){\strut{}$\lambda(\rho)$}}}%
      \put(2860,154){\makebox(0,0){\strut{}$\rho/r_\textrm{vdW}$}}%
    }%
    \gplgaddtomacro\gplfronttext{%
      \csname LTb\endcsname%
      \put(1676,3090){\makebox(0,0)[l]{\strut{}One-channel, $r_e=0$}}%
      \csname LTb\endcsname%
      \put(1676,2870){\makebox(0,0)[l]{\strut{}Two-channel, $r_e=-0.1$}}%
      \csname LTb\endcsname%
      \put(1676,2650){\makebox(0,0)[l]{\strut{}Two-channel, $r_e=-1$}}%
      \csname LTb\endcsname%
      \put(1676,2430){\makebox(0,0)[l]{\strut{}Two-channel, $r_e=-5$}}%
    }%
    \gplbacktext
    \put(0,0){\includegraphics{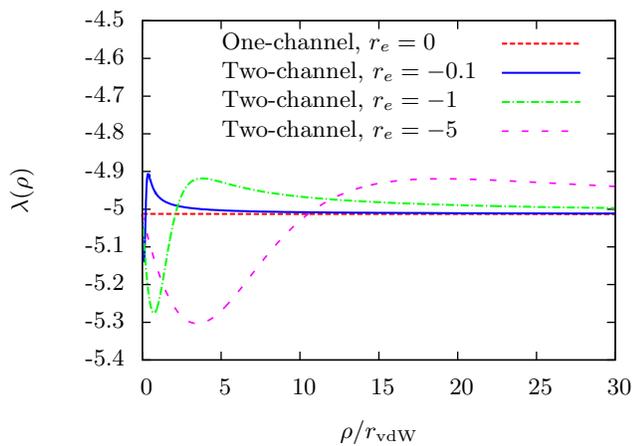}}
    \gplfronttext
  \end{picture}%
\endgroup
\caption{(Color online) The lowest adiabatic potentials multiplied by $\rho^2$
as functions of hyperradius, $\rho$, for the zero range one-channel model
and three different effective ranges for the two-channel model.}
\label{fig3}
\end{figure}

\begin{figure}[ht!]
\centering
\begingroup
  \makeatletter
  \providecommand\color[2][]{%
    \GenericError{(gnuplot) \space\space\space\@spaces}{%
      Package color not loaded in conjunction with
      terminal option `colourtext'%
    }{See the gnuplot documentation for explanation.%
    }{Either use 'blacktext' in gnuplot or load the package
      color.sty in LaTeX.}%
    \renewcommand\color[2][]{}%
  }%
  \providecommand\includegraphics[2][]{%
    \GenericError{(gnuplot) \space\space\space\@spaces}{%
      Package graphicx or graphics not loaded%
    }{See the gnuplot documentation for explanation.%
    }{The gnuplot epslatex terminal needs graphicx.sty or graphics.sty.}%
    \renewcommand\includegraphics[2][]{}%
  }%
  \providecommand\rotatebox[2]{#2}%
  \@ifundefined{ifGPcolor}{%
    \newif\ifGPcolor
    \GPcolorfalse
  }{}%
  \@ifundefined{ifGPblacktext}{%
    \newif\ifGPblacktext
    \GPblacktexttrue
  }{}%
  \let\gplgaddtomacro\g@addto@macro
  \gdef\gplbacktext{}%
  \gdef\gplfronttext{}%
  \makeatother
  \ifGPblacktext
    \def\colorrgb#1{}%
    \def\colorgray#1{}%
  \else
    \ifGPcolor
      \def\colorrgb#1{\color[rgb]{#1}}%
      \def\colorgray#1{\color[gray]{#1}}%
      \expandafter\def\csname LTw\endcsname{\color{white}}%
      \expandafter\def\csname LTb\endcsname{\color{black}}%
      \expandafter\def\csname LTa\endcsname{\color{black}}%
      \expandafter\def\csname LT0\endcsname{\color[rgb]{1,0,0}}%
      \expandafter\def\csname LT1\endcsname{\color[rgb]{0,1,0}}%
      \expandafter\def\csname LT2\endcsname{\color[rgb]{0,0,1}}%
      \expandafter\def\csname LT3\endcsname{\color[rgb]{1,0,1}}%
      \expandafter\def\csname LT4\endcsname{\color[rgb]{0,1,1}}%
      \expandafter\def\csname LT5\endcsname{\color[rgb]{1,1,0}}%
      \expandafter\def\csname LT6\endcsname{\color[rgb]{0,0,0}}%
      \expandafter\def\csname LT7\endcsname{\color[rgb]{1,0.3,0}}%
      \expandafter\def\csname LT8\endcsname{\color[rgb]{0.5,0.5,0.5}}%
    \else
      \def\colorrgb#1{\color{black}}%
      \def\colorgray#1{\color[gray]{#1}}%
      \expandafter\def\csname LTw\endcsname{\color{white}}%
      \expandafter\def\csname LTb\endcsname{\color{black}}%
      \expandafter\def\csname LTa\endcsname{\color{black}}%
      \expandafter\def\csname LT0\endcsname{\color{black}}%
      \expandafter\def\csname LT1\endcsname{\color{black}}%
      \expandafter\def\csname LT2\endcsname{\color{black}}%
      \expandafter\def\csname LT3\endcsname{\color{black}}%
      \expandafter\def\csname LT4\endcsname{\color{black}}%
      \expandafter\def\csname LT5\endcsname{\color{black}}%
      \expandafter\def\csname LT6\endcsname{\color{black}}%
      \expandafter\def\csname LT7\endcsname{\color{black}}%
      \expandafter\def\csname LT8\endcsname{\color{black}}%
    \fi
  \fi
  \setlength{\unitlength}{0.0500bp}%
  \begin{picture}(5040.00,3528.00)%
    \gplgaddtomacro\gplbacktext{%
      \csname LTb\endcsname%
      \put(946,704){\makebox(0,0)[r]{\strut{}-1.2}}%
      \put(946,1070){\makebox(0,0)[r]{\strut{}-1}}%
      \put(946,1435){\makebox(0,0)[r]{\strut{}-0.8}}%
      \put(946,1801){\makebox(0,0)[r]{\strut{}-0.6}}%
      \put(946,2166){\makebox(0,0)[r]{\strut{}-0.4}}%
      \put(946,2532){\makebox(0,0)[r]{\strut{}-0.2}}%
      \put(946,2897){\makebox(0,0)[r]{\strut{} 0}}%
      \put(946,3263){\makebox(0,0)[r]{\strut{} 0.2}}%
      \put(1276,484){\makebox(0,0){\strut{} 0.8}}%
      \put(1672,484){\makebox(0,0){\strut{} 1}}%
      \put(2068,484){\makebox(0,0){\strut{} 1.2}}%
      \put(2464,484){\makebox(0,0){\strut{} 1.4}}%
      \put(2861,484){\makebox(0,0){\strut{} 1.6}}%
      \put(3257,484){\makebox(0,0){\strut{} 1.8}}%
      \put(3653,484){\makebox(0,0){\strut{} 2}}%
      \put(4049,484){\makebox(0,0){\strut{} 2.2}}%
      \put(4445,484){\makebox(0,0){\strut{} 2.4}}%
      \put(176,1983){\rotatebox{-270}{\makebox(0,0){\strut{}$V(r)/\epsilon_0$}}}%
      \put(2860,154){\makebox(0,0){\strut{}$r/r_\textrm{rmin}$}}%
    }%
    \gplgaddtomacro\gplfronttext{%
      \csname LTb\endcsname%
      \put(3656,1317){\makebox(0,0)[r]{\strut{}Van der Waals}}%
      \csname LTb\endcsname%
      \put(3656,1097){\makebox(0,0)[r]{\strut{}Morse}}%
      \csname LTb\endcsname%
      \put(3656,877){\makebox(0,0)[r]{\strut{}Lennard-Jones}}%
    }%
    \gplbacktext
    \put(0,0){\includegraphics{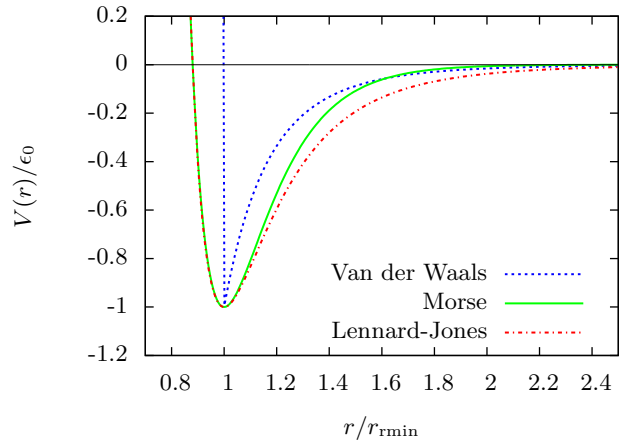}}
    \gplfronttext
  \end{picture}%
\endgroup
\caption{(Color online) Typical two-body neutral atom-atom potentials, including 
the hardcore van der Waals potential of \eq{eq:8}, the Lennard-Jones 
potential, $V_{LJ}(r)=\tfrac{C_{10}}{r^{10}}-\tfrac{C_6}{r^6}$, and the Morse 
potential, $V_M(r) = 4\epsilon_0(e^{-2\alpha(r-r_0)}-e^{-\alpha(r-r_0)})$. Here
$C_6$, $C_{10}$, and $\alpha$ are parameters. They have all been fitted to have 
the same overall energy, $\epsilon_0$, at the radius of the potential 
minimum, $r_\textrm{min}$.}
\label{fig4}
\end{figure}

\paragraph*{Connection to two-body potential.}
The zero-range model studied above does not carry information about the
van der Waals length as it stands. However, the three-body parameter
or cut-off, $\rho_c$, has a physical meaning as it 
provides a hard-core repulsion in hyperspherical three-body coordinates.
To connect the formalism to experimental data, it is therefore necessary
to find a relation between the two-body atomic physics and $\rho_c$. Below
we consider different two-body interaction models and derive analytical 
formulas for the behavior of $a^{(-)}$ as the two-body interaction
parameters are varied.

In Fig.~\ref{fig4} we plot Morse, Lennard-Jones, and a van der Waals
potential with a hard-core at $r_c$. We first focus on 
the van der Waals plus hard-core model, 
\begin{equation}
V(r)=-\frac{C_6}{r^6},\quad r>r_c
	\label{eq:8}
\end{equation}
where $r_\textrm{vdW}=(\tfrac{mC_6}{\hbar^2})^{1/4}$.
In order to relate the behavior of this potential to the physics of a Feshbach resonance, 
we use the formula \cite{PhysRevA.48.1993,PhysRevA.59.1998}
$a=\bar{a}(1-\tan\Phi)$ with $\Phi=\tfrac{2\vdW^2}{r_c^2}-\tfrac{3\pi}{8}$
where $a$ diverges when $\Phi=(n+\tfrac12)\pi$ for integer $n$, which 
counts the number of $s$-wave bound states accomodated by the 
potential. Thus
\begin{equation}
	n=\frac{2}{\pi}\left(\frac{r_\textrm{vdW}}{r_c}\right)^2-\frac{7}{8}
	\label{eq:15}
\end{equation}
rounded to highest nearby integer. 

\begin{figure}[ht!]
\centering
\begingroup
  \makeatletter
  \providecommand\color[2][]{%
    \GenericError{(gnuplot) \space\space\space\@spaces}{%
      Package color not loaded in conjunction with
      terminal option `colourtext'%
    }{See the gnuplot documentation for explanation.%
    }{Either use 'blacktext' in gnuplot or load the package
      color.sty in LaTeX.}%
    \renewcommand\color[2][]{}%
  }%
  \providecommand\includegraphics[2][]{%
    \GenericError{(gnuplot) \space\space\space\@spaces}{%
      Package graphicx or graphics not loaded%
    }{See the gnuplot documentation for explanation.%
    }{The gnuplot epslatex terminal needs graphicx.sty or graphics.sty.}%
    \renewcommand\includegraphics[2][]{}%
  }%
  \providecommand\rotatebox[2]{#2}%
  \@ifundefined{ifGPcolor}{%
    \newif\ifGPcolor
    \GPcolortrue
  }{}%
  \@ifundefined{ifGPblacktext}{%
    \newif\ifGPblacktext
    \GPblacktexttrue
  }{}%
  \let\gplgaddtomacro\g@addto@macro
  \gdef\gplbacktext{}%
  \gdef\gplfronttext{}%
  \makeatother
  \ifGPblacktext
    \def\colorrgb#1{}%
    \def\colorgray#1{}%
  \else
    \ifGPcolor
      \def\colorrgb#1{\color[rgb]{#1}}%
      \def\colorgray#1{\color[gray]{#1}}%
      \expandafter\def\csname LTw\endcsname{\color{white}}%
      \expandafter\def\csname LTb\endcsname{\color{black}}%
      \expandafter\def\csname LTa\endcsname{\color{black}}%
      \expandafter\def\csname LT0\endcsname{\color[rgb]{1,0,0}}%
      \expandafter\def\csname LT1\endcsname{\color[rgb]{0,1,0}}%
      \expandafter\def\csname LT2\endcsname{\color[rgb]{0,0,1}}%
      \expandafter\def\csname LT3\endcsname{\color[rgb]{1,0,1}}%
      \expandafter\def\csname LT4\endcsname{\color[rgb]{0,1,1}}%
      \expandafter\def\csname LT5\endcsname{\color[rgb]{1,1,0}}%
      \expandafter\def\csname LT6\endcsname{\color[rgb]{0,0,0}}%
      \expandafter\def\csname LT7\endcsname{\color[rgb]{1,0.3,0}}%
      \expandafter\def\csname LT8\endcsname{\color[rgb]{0.5,0.5,0.5}}%
    \else
      \def\colorrgb#1{\color{black}}%
      \def\colorgray#1{\color[gray]{#1}}%
      \expandafter\def\csname LTw\endcsname{\color{white}}%
      \expandafter\def\csname LTb\endcsname{\color{black}}%
      \expandafter\def\csname LTa\endcsname{\color{black}}%
      \expandafter\def\csname LT0\endcsname{\color{black}}%
      \expandafter\def\csname LT1\endcsname{\color{black}}%
      \expandafter\def\csname LT2\endcsname{\color{black}}%
      \expandafter\def\csname LT3\endcsname{\color{black}}%
      \expandafter\def\csname LT4\endcsname{\color{black}}%
      \expandafter\def\csname LT5\endcsname{\color{black}}%
      \expandafter\def\csname LT6\endcsname{\color{black}}%
      \expandafter\def\csname LT7\endcsname{\color{black}}%
      \expandafter\def\csname LT8\endcsname{\color{black}}%
    \fi
  \fi
  \setlength{\unitlength}{0.0500bp}%
  \begin{picture}(5040.00,3528.00)%
    \gplgaddtomacro\gplbacktext{%
      \csname LTb\endcsname%
      \put(814,704){\makebox(0,0)[r]{\strut{} 0}}%
      \put(814,1216){\makebox(0,0)[r]{\strut{} 5}}%
      \put(814,1728){\makebox(0,0)[r]{\strut{} 10}}%
      \put(814,2239){\makebox(0,0)[r]{\strut{} 15}}%
      \put(814,2751){\makebox(0,0)[r]{\strut{} 20}}%
      \put(814,3263){\makebox(0,0)[r]{\strut{} 25}}%
      \put(946,484){\makebox(0,0){\strut{} 0}}%
      \put(1685,484){\makebox(0,0){\strut{} 5}}%
      \put(2425,484){\makebox(0,0){\strut{} 10}}%
      \put(3164,484){\makebox(0,0){\strut{} 15}}%
      \put(3904,484){\makebox(0,0){\strut{} 20}}%
      \put(4643,484){\makebox(0,0){\strut{} 25}}%
      \put(176,1983){\rotatebox{-270}{\makebox(0,0){\strut{}$a^{(-)}/\vdW$}}}%
      \put(2794,154){\makebox(0,0){\strut{}$n$}}%
      \csname LTb\endcsname%
      \put(2425,1369){\makebox(0,0)[l]{\strut{}${}^7$Li}}%
      \put(1538,3058){\makebox(0,0)[l]{\strut{}${}^{39}$K}}%
      \put(3016,1932){\makebox(0,0)[l]{\strut{}${}^{85}$Rb}}%
      \put(3608,2035){\makebox(0,0)[l]{\strut{}${}^{133}$Cs}}%
    }%
    \gplgaddtomacro\gplfronttext{%
      \csname LTb\endcsname%
      \put(3656,3090){\makebox(0,0)[r]{\strut{}Mean value}}%
      \csname LTb\endcsname%
      \put(3656,2870){\makebox(0,0)[r]{\strut{}One-channel}}%
      \csname LTb\endcsname%
      \put(3656,2650){\makebox(0,0)[r]{\strut{}Two-channel, $\log(s_\text{res})=0$}}%
      \csname LTb\endcsname%
      \put(3656,2430){\makebox(0,0)[r]{\strut{}Two-channel, $\log(s_\text{res})=-0.7$}}%
    }%
    \gplbacktext
    \put(0,0){\includegraphics{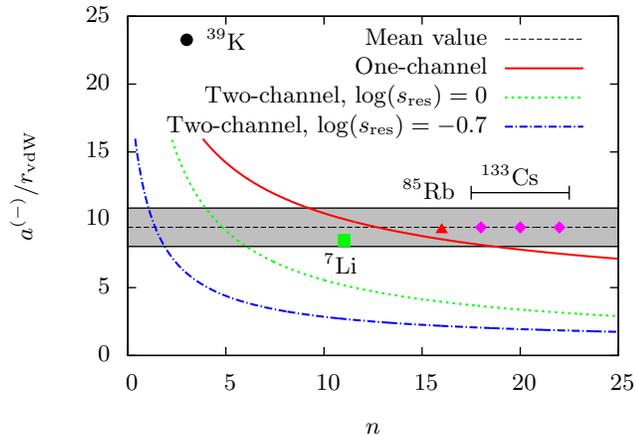}}
    \gplfronttext
  \end{picture}%
\endgroup
\caption{(Color online) Semianalytic results for $a^{(-)}$ plotted against number of bound states 
in the two-body van der Waals plus hard-core potential. 
The horizontal position of the experimental data is arbitrary. The grey 
band indicates the 15\% margin around the mean value of $\sim9.8$.}
\label{fig5}
\end{figure}

In the zero-range single-channel model $\rho_c$ and $a^{(-)}$ are related, since $\rho_c$ 
is the only length scale available. However, it is a non-trivial matter to determine
this relation within the model. Numerically, we find the linear relation
$a^{(-)} = -\delta\,\rho_c$,
with $\delta=31.756$. To cast this relation into a form that depends only on two-body
physics, we observe that the two-body hard-core at $r_c$ is also 
responsible for a three-body hard-core cut-off, more precisely $\rho_c=\sqrt{2}r_c$. 
This condition ensures that when each two-body subsystem has radius $r\geq r_c$, the 
third particle will also be outside $r_c$ with respect to the others (see 
Ref.~\cite{jen97} and the supplementary note \ref{appa}).
We note that this relation is not the same as the $\rho_c$ used in 
Ref.~\cite{chin2011}.  Using $a^{(-)} = -\delta\,\rho_c$ we obtain
\begin{equation}
	\frac{a^{(-)}}{\vdW}=-\frac{2\delta}{\sqrt{(n+\frac78)\pi}}.
	\label{eq:18}
\end{equation}
This semi-analytical expression for the threshold in terms of the number of 
bound states is one of our main results.

The relation in Eq.~\eqref{eq:18} is plotted in Fig.~\ref{fig5} along 
with the experimental data and the numerical results
obtained from the two-channel model for different value of $s_\textrm{res}$.
The one-channel model is consistent with data for $n\sim 10-20$
and fits the universal ratio of Ref.~\cite{chin2011} for $n=13$. This is also
consistent with the findings of Ref.~\cite{wang2012}, although
the data only goes to $n=10$. Actually, the 
$n^{-1/2}$ behavior that we find here seems to also appear in Ref.~\cite{wang2012}, 
where an extension to higher $n$
could confirm this prediction.

\setlength{\tabcolsep}{15pt}
\begin{table}[h]
	\centering
	\begin{tabular}{crrrrr}
		\hline\noalign{\smallskip}
		& Li & K & Rb & Cs \\
		\hline\noalign{\smallskip}
		$R_e$ [\AA] & 2.67 & 3.92 & 4.18 & 4.65\\
		$D_e$ [eV] & 1.06 & 0.52 & 0.49 & 0.45\\
		$n_{LJ}$ & 41 & 99 & 152 & 201\\
		$n_M$ & 28 & 67 & 103 & 137\\
		\noalign{\smallskip}\hline
	\end{tabular}
	\caption{Potential parameters, $R_e$ (bond length) and $D_e$ (dissociation energy) \cite{Igel-Mann}, and estimated number of bound states for 
	Lennard-Jones, $n_{LJ}$, and Morse, $n_M$, potentials.}
	\label{tab1}
\end{table}

Even more interestingly, our results for small $s_\textrm{res}$ indicate that
$|a^{(-)}|$ drops faster with $n$ than for $s_\textrm{res}\gg 1$. This
is seen in the experimental data on $^{7}$Li which is slightly below the 
$^{85}$Rb and $^{133}$Cs points, but our model seems to overestimate this
trend. Clearly, more results on narrow resonance systems are required to
address the question of effective range corrections.
We expect a lower $|a^{(-)}|$ value than for broad resonances.

\paragraph*{Two-body potential models}
Above we employed a van der Waals plus hard-core potential. However, 
as seen in Fig.~\ref{fig4}, more realistic Lennard-Jones or Morse 
potentials have a smooth behavior of the inner barrier. This implies
only minor quantitative corrections that nevertheless deserve to 
be addressed along with the number of bound states expected in a real
alkali dimer.

The number of $s$-wave bound states in the Lennard-Jones (LJ) and Morse (M) potentials
can be estimated analytically and yields \cite{MahanLapp}
$n_{LJ}=0.361\sqrt\beta-\tfrac{5}{8}$ and
$n_{M}=0.245\sqrt\beta-\tfrac{1}{2}$,
where $\beta=\tfrac{m r_\textrm{min}^{2}\epsilon_0}{2\hbar^2}$ with $r_\textrm{min}$ 
the radius at which the potential takes its minimal value,
$\epsilon_0$. For comparison, the expression in 
Eq.~\eqref{eq:15} can be written
$n=0.225\sqrt{\beta}-\tfrac{7}{8}$, 
with $r_\textrm{min}\leftrightarrow r_c$ in $\beta$. 
The similarity of these expressions makes it clear that the 
behavior seen in Fig.~\ref{fig5} is generic and does not 
depend on our choice of two-body potential. The difference
in constant provides only a minor quantitative change in the numbers. 

An important question, however, remains about the 
number of bound states in a real alkali
dimer system. 
Using the molecular 
dissociation energy, $\epsilon_0=D_e$, and the bond length, $r_\textrm{min}=R_e$, 
of Ref.~\cite{Igel-Mann},
we list the numbers for Li, K, Rb, and Cs dimers in Tab.~\ref{tab1}.
The estimated number of bound states is outside
the axis in Fig.~\ref{fig5} and also much beyond the results shown in 
Ref.~\cite{wang2012}. The agreement with theory at a rather limited number 
of bound states ($n\sim 10-20$) is then quite surprising. 

A number of important observations can be made. First, the decrease of 
$|a^{(-)}|$ with $n$ is weak, and a shift of the length scale in Fig.~\ref{fig5}
would therefore place the one-channel model within the experimental range 
for larger $n$ and it would stay within the 15\% deviation from the mean
for a larger interval (since the slope at larger $n$ decreases even faster).
Second, the experimental data might indicate that only a certain number
of bound states play an active role. Equivalently, even if the two-body
potential is very deep, only the upper part of the two-body potential and 
the bound states closest to threshold set the scale of the 
three-body problem.
This appears to be very reasonable since we are
considering universal Efimov trimers here and not strongly bound
three-body states. Third, the case of small $s_\textrm{res}$ 
has $|a^{(-)}|\propto n^{-r}$ with $r>1/2$ as seen in Fig.~\ref{fig5}. 
This implies that narrow resonance systems should be even more insensitive
to $n$ beyond a certain lower limit. 

We can give a quantitative argument for the lack of sensitivity to the many 
deeper bound states in a van der Waals potential. The 
number of bound states, $n(E)$ as a function of energy, $E$, counted from 
the $E=0$ threshold and down within the WKB approximation is $n(E)=n_0(|E|/E_\textrm{vdW})^{1/3}$,
where $n_0$ is the total number (given in Tab.~\ref{tab1} for different potentials)
and $E_\textrm{vdW}=\hbar^2/mr_\textrm{vdW}^{2}>0$ is the depth of the potential \cite{WKB}. 
For the $s_\textrm{res}\gg 1$
cases ($^{85}$Rb and $^{133}$Cs), $n(E)/n_0\sim 0.10-0.20$ which implies
$|E|/E_\textrm{vdW}\sim 0.001-0.01$. Numerically we find a three-body energy on resonance 
$E_3=0.006E_\textrm{vdW}$ (using $\rho_c=0.58r_\textrm{vdW}$). However, universality
relates $E_3=\hbar^2\kappa^2/m$ and $a^{(-)}\kappa\sim -1.51$ \cite{efi70,braaten2006}. 
The energy scale at the continuum threshold is given by 
$a^{(-)}$ through $|E|\sim \hbar^2/m(a^{(-)})^{2}=0.003E_\textrm{vdW}$, in agreement
with the interval above. In the case of $^{7}$Li, $E_3$ is similar but this is
compensated by a smaller $n_0$ so this case can also be explained.
For heavier alkali atoms with narrow resonance, our two-channel results predict a 
smaller $a^{(-)}/r_\textrm{vdW}$ than 9.8, which is a good experimental test
of our theory.

\paragraph*{Conclusions.}
We derive an analytical formula that connects $a^{(-)}$ 
and the number of $s$-wave bound states, $n$, in the two-body potential which 
agrees with recent experiments $n\sim 10-20$. While alkali atoms typically have
larger $n$, we argue quantitatively that only a subset determine the 
properties of Efimov three-body states.
We find non-monotonic behavior of $a^{(-)}$ with
increasing effective range which demonstrates that systems with 
narrow Feshbach resonances will provide both qualitative and
quantitative understanding of universality 
and the origin of the three-body parameter. This means that 
different atomic species which typically have narrow resonances 
and a denser spectrum would be very helpful.

\paragraph*{Acknowledgments} 
We thank Chris Greene, Georg Bruun, and Frank Jensen for useful discussions. This work is supported 
by the Danish Council for Independent Research -- Natural Sciences (DFF-FNU).

\appendix

\section{Determination of the hyperradial boundary condition}\label{appa}
The atom-atom two-body potential has a steep repulsive inner core which we will assume for the 
moment can be representated by a hard wall, i.e. an infinite potential for radii $r<r_c$. In this
case the boundary condition is simply that the two-body wave function must be zero at $r_c$. This 
must then be translated in the three-body problem where it implies that the total wave function 
must be zero whenever any of the relative distances between two out of the three atoms is less than
or equal to $r_c$. Any penetration of the wave function into the wall would cost an infinite
amount of energy and is thus forbidden.

\begin{figure}[t]
\includegraphics[scale=0.35]{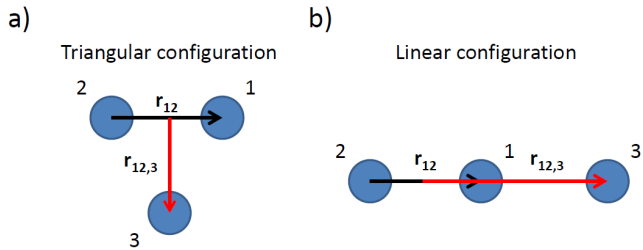} 
\caption{(Color online) Schematic of the a) triangular and b) linear configurations for an equal mass 
three-body system. $\bm r_{12}$ and $\bm r_{12,3}$ indicate the two Jacobi relative vectors.}
\label{fig6}
\end{figure}

A neat and elegant way to obtain a condition on $\rho$ is the following. As is easily shown the 
hyperradius fulfills
\begin{align}
\rho^2=\frac{1}{3}\sum_{i<k}\left(\bm r_i-\bm r_k\right)^2=\frac{1}{2}\bm r^{2}_{12}+\frac{1}{3}\bm r^{2}_{12,3},
\label{rho}
\end{align}
where $\bm r_{12}=\bm r_1-\bm r_2$ and $\bm r_{12,3}=\bm r_3-(\bm r_1+\bm r_2)/2$ are respectively the relative 
vector from particles 2 to 1 and the relative vector from the center of mass of 1 and 2 to particle 3 (see
Fig~\ref{fig6}).

Now consider the triangular configuration shown in a) of Fig.~\ref{fig6}. From the 
formula in Eq.~\eqref{rho}, it is clear that when $r_{12}=r_c$ this setup also has $\rho=r_c$. 
However, the linear configuration of b) in Fig.~\ref{fig6} with $\rho=r_c$ will then have 
two atoms that are within the hard-core radius and yield an infinite contribution to the 
potential energy (this happens f.x. when $|\bm x|=r_c$ and $\bm y=0$ putting atom 3 midway 
between 1 and 2).

Consider instead the linear configuration with $r_{12}=r_c$ and impose the requirement
$r_{23}\geq r_c$, where $\bm r_{23}=\bm r_2-\bm r_3$. Since we have $\bm r_{12,3}=\bm r_{23}+\bm r_{12}/2$ we get
\begin{align}
\rho^2\geq r_{c}^{2}\left( \frac{1}{2} + \frac{2}{3}[1+\frac{1}{2}]^2\right)=2r_{c}^{2}.
\end{align}
We thus see that the condition $\rho>\sqrt{2}r_c$ ensures that both triangular and linear
configurations are outside the hard-core regions. Since these configurations are extremal, 
the condition implies that no regions with infinite potential are sampled by the hyperradial 
three-body wave function.

The rigourous formal argument for the validity of the relation $\rho_c=\sqrt{2}r_c$ 
using the hyperspherical approach can be found in Ref.~\cite{jen97}, 
where the relation is derived using a square well potential. The asymptotic region is precisely
$\rho>\sqrt{2}r_c$ as found above. Here we assumed a hard-core potential for simplicity which 
gives the factor of $\sqrt{2}$. For a real atom-atom potential, the hard-core is slightly 
softer (typically of the $1/r^{12}$) which may lead to a minor change in the factor.

The hyperspherical approach and the use of the lowest hyperradial potential to describe 
Efimov states is a well-known and accurate procedure. Additionally, as we have shown above, a
hard-core repulsion in the two-body potential naturally generates a hyperspherical cut-off
radius, $\rho_c=\sqrt{2}r_c$. This constitutes a clear and simple explanation of the 
repulsive core found in the numerical calculations in Ref.~\cite{wang2012}, and our results
demonstrate that a simple model can accurately describe those numerically involved findings.
We thus see how the value $\sqrt{2}r_c$ for the three-body cut-off parameter naturally 
derives from the hard-core behavior of the corresponding two-body potential.


\begin{thebibliography}{99}
\bibitem{WKB} The number of bound states, $n(E)$, can be found within the WKB formalism by considering the 
integral $\int_{r_i}^{r_o}\sqrt{E-V(r)}dr=\pi \hbar(n(E)+1/4)$ which is a measure of the number of nodes
in the wave function with energy $E<0$. Using $E=0$ one obtains the total number, $n_0$. The van der Waals
potential does not yield analytical results, but we numerically find that $n(E)\propto |E|^{1/3}$, which 
when normalized yields the relation in the text.
\end{thebibliography}
\end{document}